\documentclass[a4paper]{article}

\usepackage{INTERSPEECH2019}
\usepackage{array}
\usepackage{graphicx}
\usepackage{float}
\usepackage{subfigure}
\usepackage{enumerate}
\usepackage{multirow}
\usepackage{xcolor}

\title{LSTM based Similarity Measurement with Spectral Clustering \\for Speaker Diarization}
\name{Qingjian Lin$^{1,2}$, 
      Ruiqing Yin$^3$,
      Ming Li$^1$,
      Herv\'e Bredin$^3$,
      Claude Barras$^3$}
\address{$^1$Data Science Research Center, Duke Kunshan University, Kunshan, China\\ 
         $^2$School of Electronics and Information Technology, Sun Yat-sen University, Guangzhou, China,
         $^3$LIMSI, CNRS, Univ. Paris-Sud, Universit\'{e} Paris-Saclay, Orsay, France}
\email{ming.li369@dukekunshan.edu.cn}

\begin{document}

\maketitle
\begin{abstract}
  More and more neural network approaches have achieved considerable improvement upon submodules of speaker diarization system, including speaker change detection and segment-wise speaker embedding extraction. Still, in the clustering stage, traditional algorithms like probabilistic linear discriminant analysis (PLDA) are widely used for scoring the similarity between two speech segments. In this paper, we propose a supervised method to measure the similarity matrix between all segments of an audio recording with sequential bidirectional long short-term memory networks (Bi-LSTM). Spectral clustering is applied on top of the similarity matrix to further improve the performance. Experimental results show that our system significantly outperforms the state-of-the-art methods and achieves a diarization error rate of 6.63\% on the NIST SRE 2000 CALLHOME database.
\end{abstract}


\section{Introduction}

\label{sec:intro}
Speaker diarization is the task of determining \emph{``who speaks when''} in an audio file that usually contains an unknown number of speakers with variable speech duration~\cite{tranter2006overview,anguera2012speaker}. 

Diarization systems are usually made of multiple submodules. First, a voice activity detector (VAD)~\cite{price2018low} removes non-speech regions from the audio input. Then, speech regions are split into multiple speaker-homogeneous segments either with a speaker change detector (SCD)~\cite{hruz2017convolutional,yin2017speaker} or based on uniform segmentation~\cite{wang2018speaker}. These segments are mapped into a fixed-dimensional feature space by speaker embedding systems such as i-vector~\cite{shum2013unsupervised,sell2014speaker}, x-vector~\cite{snyder2018x,garcia2017speaker}, or penultimate layer output from various end-to-end speaker verification methods~\cite{Cai2018Exploring,Cai2018Analysis}. Next, pairwise similarity measurement techniques like cosine distance and PLDA~\cite{sell2014speaker, prince2007probabilistic} compute the similarity matrix between segments. Finally, agglomerative hierarchical clustering (AHC)~\cite{meignier2010lium}, spectral clustering~\cite{wang2018speaker} or affinity propagation~\cite{yin2018neural} are applied on top of the similarity matrix to obtain the final diarization results. 

While the performance of speech recognition and speaker verification systems has improved dramatically thanks to deep learning approaches, most speaker diarization systems have not yet taken full advantage of these techniques. One reason is that speaker diarization labels are ambiguous: both \emph{``111223''} and \emph{``222113''} sequences can be equally correct sequences of labels for the same audio input file. Because it is usually addressed as an unsupervised task, the clustering step makes it difficult to design fully supervised diarization system. Zhang~\cite{zhang2018fully} did propose the UIS-RNN model for clustering and improved the performance, but UIS-RNN is essentially a mixture of LSTMs and parametric models, and relies heavily on the effectiveness of the speaker embedding front-end.

In this work, we propose to use Bi-LSTM in place of PLDA to model the similarity between any arbitrary two segments. Since PLDA scores similarity between two segments in a pairwise and independent manner, it completely ignores the sequential order of speech segments. However, conversations between several speakers are usually highly structured, and turn-taking behaviors are not randomly distributed over time. In~\cite{wisniewksi2017combining}, structured prediction is applied for online speaker diarization, but only the structural information from the forward direction is considered. We propose to use both forward and backward segments to overcome such limitations and enhance the performance of similarity measurement. Besides, we use spectral clustering on top of the similarity matrix to obtain the final results.

The rest of this paper is organized as follows. The next section presents the general diarization system overview including speaker embedding extraction, similarity measurement and clustering algorithms. Section 3 describes the details of our Bi-LSTM based similarity measurement module. Experimental results and discussions are presented in Section 4 and conclusions are drawn in Section 5.

\section{System overview}
\begin{figure}[t]
  \centering
  \includegraphics[width=\linewidth]{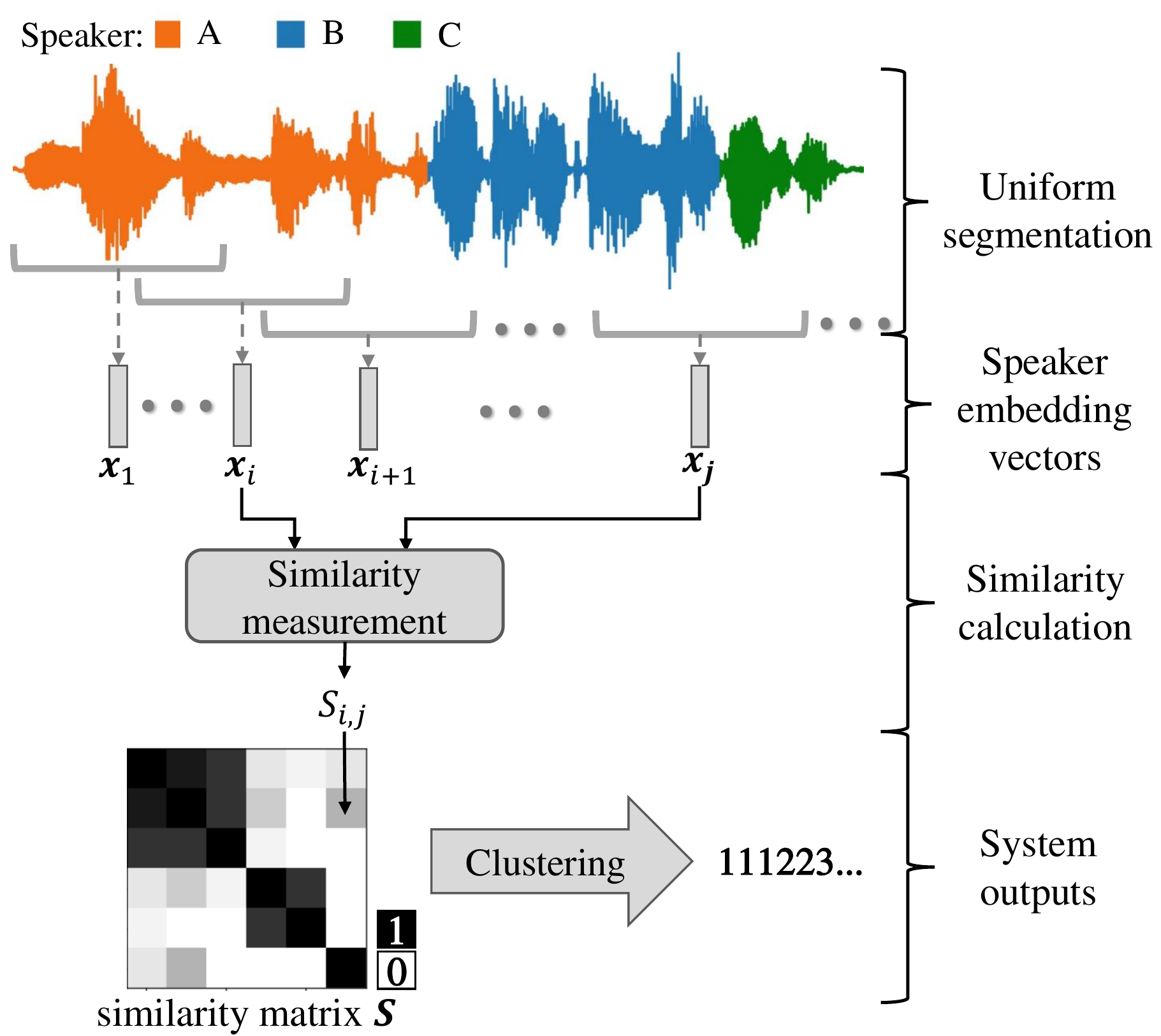}
  \caption{System framework of speaker diarization.}
  \label{fig:framework}
\end{figure}

In this paper, an oracle VAD is employed to remove non-speech regions in audios. An overview of our system framework is shown in Figure~\ref{fig:framework}. First, we employ uniform segmentation and extract $d$-dimensional speaker embedding vectors $\boldsymbol{x}_1, \boldsymbol{x}_2, ... \boldsymbol{x}_n$ from assumedly speaker-homogeneous segments with pre-trained speaker embedding models. Second, a similarity measurement algorithm computes the score $S_{ij}$ between every embedding vector pair $(\boldsymbol{x}_i, \boldsymbol{x}_j), i, j \in \{1, 2,...n\}$ and form a square similarity matrix $\boldsymbol{S}$. Finally, we perform clustering among all segments based on $\boldsymbol{S}$. Related algorithms are briefly introduced in this section.

\subsection{Speaker embedding vectors}
\subsubsection{i-vector}
i-vector is one of the most widely used speaker embedding vector. Essential components in the i-vector system include the universal background model (UBM) and the total variability space $T$. For a pre-trained system, mel-frequency cepstral coefficients (MFCCs) are extracted from input audios and used for adapting the UBM into speaker-specific gaussian mixture models (GMMs). From speaker-specific GMMs, corresponding supervectors can be computed and then projected onto subspace~$T$ as i-vectors. 

\subsubsection{x-vector}
x-vector is an approach based on deep neural networks that has demonstrated excellent performance in speaker verification. MFCCs are extracted and fed into a time-delay neural network (TDNN) for supervised learning. In the TDNN architecture, a time-pooling module transforms multiple frame-level features to a single segment-level embedding, followed by fully connected layers. The output of the penultimate linear layer is called the x-vector.

\subsection{Similarity Measurement: PLDA}

PLDA has been used successfully to measure the similarity between two segment~\cite{sell2014speaker}. Given a pre-trained model, the similarity score between segments $i$ and $j$ can be directly calculated by hypothesis testing:
$$S_{ij} =f_{\text{PLDA}}(\boldsymbol{x}_i,\boldsymbol{x}_j).$$
$S_{ij}$ describes how similar $\boldsymbol{x}_i$ and $\boldsymbol{x}_j$ are. Ideally we hope $S_{ij} = 1$ if segment $i$ and $j$ are from the same speaker, and $S_{ij} = 0$ otherwise. However, as a hypothesis testing based method, PLDA generates either negative or positive similarity scores, which raises problems in specific clustering backends like spectral clustering. For convenience, we normalize PLDA scores by a logistic function:
% Hervé: I removed the notations L, x_0, k as they were not used nor discussed anywhere else. 
$$ g(x) = \frac{1}{1+e^{-5 x}},$$

Although PLDA performs well in speaker verification tasks~\cite{kenny2013plda}, it only performs pairwise comparisons and therefore ignores the temporal structure of conversations when estimating the similarity matrix.

\subsection{Clustering backend}
\subsubsection{Agglomerative Hierarchical Clustering (AHC)}
Agglomerative Hierarchical Clustering is presented as a binary-tree building process~\cite{gowda1978agglomerative}. Segments are initialized as singleton clusters. In each iteration, clusters with the highest pairwise similarity are merged until the similarity score between any two clusters is below a given threshold~$\alpha$.

\subsubsection{Spectral Clustering (SC)}
Spectral clustering is a graph-based clustering algorithm~\cite{von2007tutorial}. Given the similarity matrix $\boldsymbol{S}$, it considers $S_{ij}$ as the weight of the edge between nodes $i$ and $j$ in an undirected graph. By removing weak edges with small weights, spectral clustering divides the original graph into subgraphs. As described in~\cite{von2007tutorial}, spectral clustering consists of the following steps:
\begin{enumerate}[a)]
\setlength{\itemsep}{1pt}
\setlength{\parsep}{0pt}
\setlength{\parskip}{0pt}
\item Construct $\boldsymbol{S}$ and set diagonal elements to 0.
\item Compute Laplacian matrix $\boldsymbol{L}$ and perform normalization:
   $$\boldsymbol{L} = \boldsymbol{D} - \boldsymbol{S}$$
   $$\boldsymbol{L}_{\text{norm}} = \boldsymbol{D}^{-1}\boldsymbol{L}$$
   where $\boldsymbol{D}$ is a diagonal matrix and $D_i = \sum_{j=1}^{n}S_{ij}$.
\item Compute eigenvalues and eigenvectors of $\boldsymbol{L}_{\text{norm}}$.
\item Compute the number of clusters $k$. One property of $\boldsymbol{L}_{\text{norm}}$ indicates that the number of clusters in the graph equals algebraic multiplicity of the 0 eigenvalue. In our implementation, we set a threshold $\beta$ and count the number of eigenvalues below $\beta$ as $k$.
\item Take the $k$ smallest eigenvalues $\lambda_{1},\lambda_{2},...\lambda_{k}$ and corresponding eigenvectors $\boldsymbol{p}_1, \boldsymbol{p}_2,...\boldsymbol{p}_k$ of $\boldsymbol{L}_{\text{norm}}$ to construct matrix $\boldsymbol{P}\in \mathbb{R}^{n\times k}$ using $\boldsymbol{p}_1,\boldsymbol{p}_2,...\boldsymbol{p}_k$ as columns.
\item Cluster row vectors $\boldsymbol{y}_1, \boldsymbol{y}_2,...\boldsymbol{y}_n$ of $\boldsymbol{P}$ using the k-means algorithm. 
% Hervé: I removed this because this notations (C and A) are never used elsewhere and don't help understanding (quite the opposite actually)
%Let's denote the k-means clustering results as  $C_{1},C_{2},...C_{k}$, then the final output clusters $A_{1},A_{2},...A_{n}$ satisfy $A_{i}=\{j\vert \boldsymbol{y}_j\in C_{i}\}$.
\end{enumerate}

\section{Bi-LSTM based scoring}
\subsection{Bi-LSTM similarity measurement}
In a reference similarity matrix $\boldsymbol{S}$, the values are all zeros and ones, representing whether each segment pair is from the same speaker or not. Besides, the similarity matrix is robust against speaker index changes or flipping. Therefore, we utilize $\boldsymbol{S}$ as the label of the entire speaker embedding sequence $\boldsymbol{x}$ for supervised diarization learning. 

\begin{figure*}[t]
  \centering
  \includegraphics[width=0.9\textwidth]{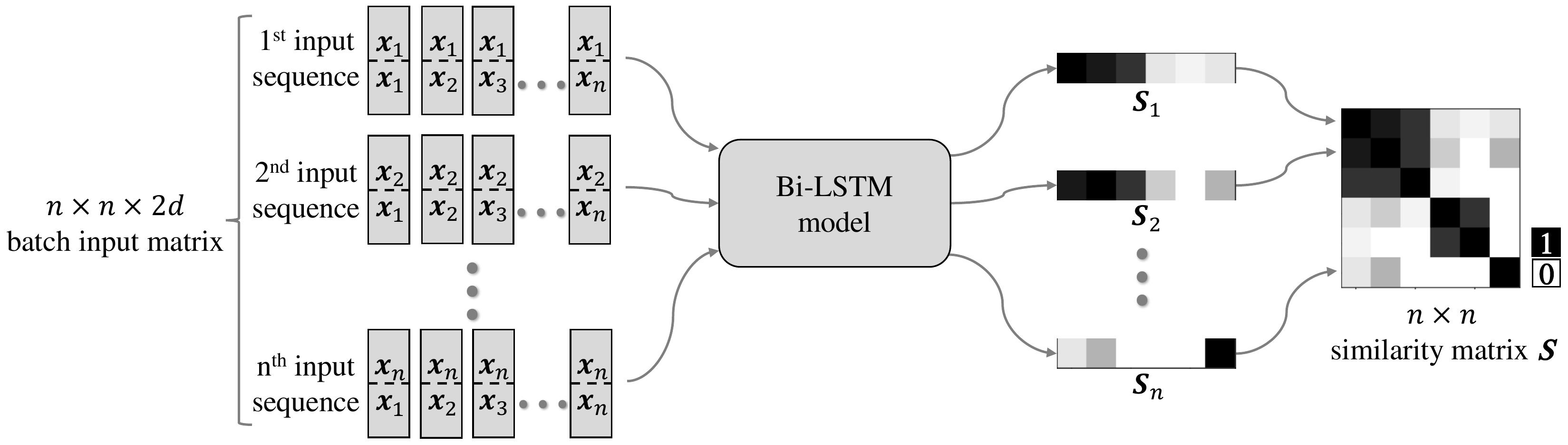}
  \caption{Bi-LSTM workflows in a batch.}
  \label{fig:model}
\end{figure*}

We propose to predict each row of the similarity matrix~$\boldsymbol{S}$ with a stacked Bi-LSTM using the binary cross entropy (BCE) loss function. Speaker embedding vectors $\boldsymbol{x}_i$ and $\boldsymbol{x}_j$ are concatenated as the $2d$-dimensional input $[\boldsymbol{x}_i^T, \boldsymbol{x}_j^T]^T$, where the corresponding output is $S_{ij}$. Since LSTMs deal with sequential data, the problem can be expressed as follows: 
$$\boldsymbol{S}_i = [S_{i1}, S_{i2}, ... S_{in}] = f_{\text{LSTM}}\left(
  \begin{bmatrix} \boldsymbol{x}_i\\ \boldsymbol{x}_1 \end{bmatrix},
  \begin{bmatrix} \boldsymbol{x}_i\\ \boldsymbol{x}_2 \end{bmatrix},\cdots
  \begin{bmatrix} \boldsymbol{x}_i\\ \boldsymbol{x}_n \end{bmatrix}\right).$$
$\boldsymbol{S}_i$ represents the $i^{th}$ row of the similarity matrix $\boldsymbol{S}$ as well as the $i^{th}$ sequence outputs in a batch. As depicted in Figure~\ref{fig:model}, there are $n$ sequences in a batch and all $n$ outputs stack row-wise as a complete similarity matrix $\boldsymbol{S}$.

Practically, since the length of the input audio is variable, in some cases the $n\times n$ matrix $\boldsymbol{S}$ can be very large (especially with uniform segmentation on long recordings). This is problematic for two reasons. First, training requires GPUs with a lot of memory. Second, it is unsure how LSTMs can handle and generalize to very long sequences. If we process the entire $n$ segments in an $m$-segment $(m<n)$ sliding window manner, the size of input and label vectors is fixed, which could help training the neural network. However, such a system eventually generates a diagonal block similarity matrix. Since part of information in the matrix is lost, it easily fails to track different speakers among different windows. For example, when the sliding window uses a step of $m-1$ segments and both $(\boldsymbol{x}_1, \boldsymbol{x}_m)$ and $(\boldsymbol{x}_m, \boldsymbol{x}_{2m-1})$ pairs are considered dissimilar, there is no evidence whether $\boldsymbol{x}_1$ and $\boldsymbol{x}_{2m-1}$ come from the same speaker due to the limited length of the window. 

\begin{figure}[!htb]
  \centering
  \includegraphics[width=\linewidth]{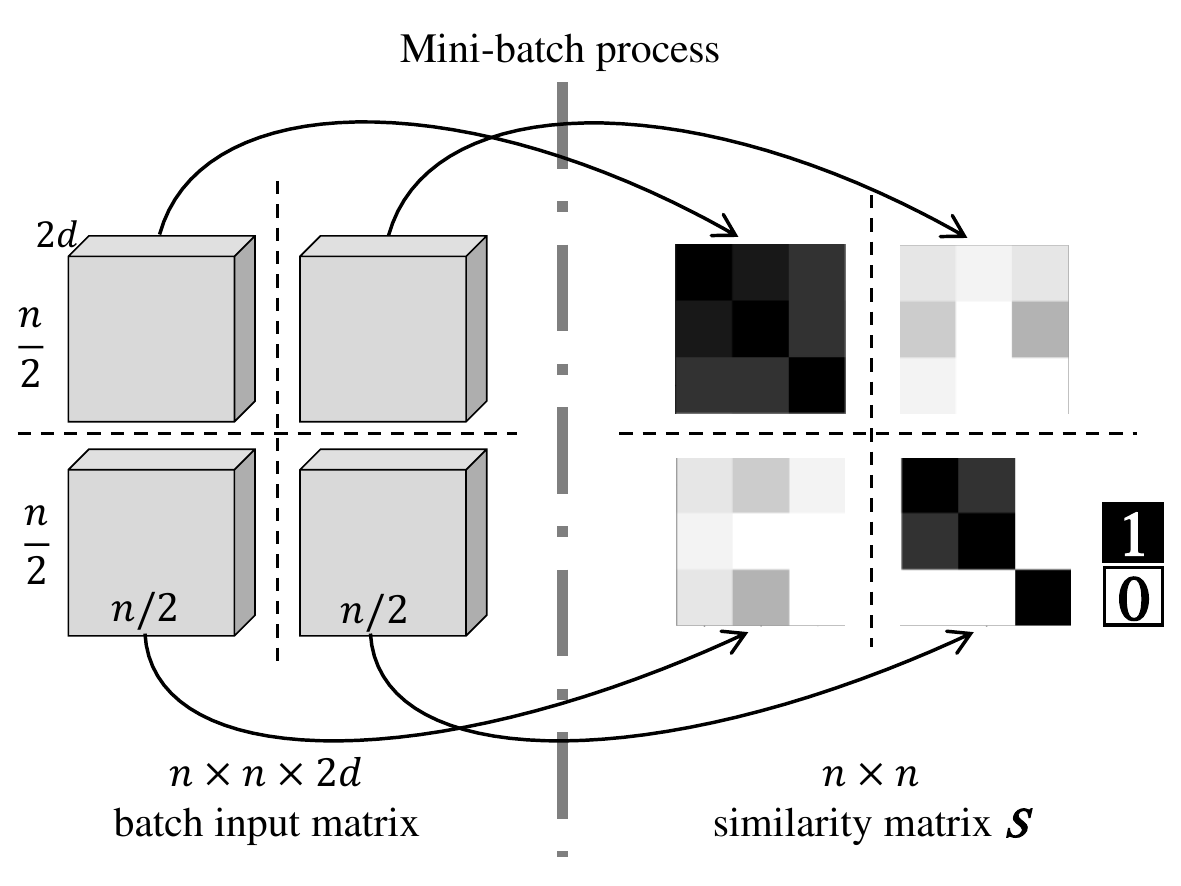}
  \caption{A large matrix is partitioned into multiple sub-matrices, processed as mini-batches.}
  \label{fig:blocks}
\end{figure}

In this case, our solution is to partition the similarity matrix into small sub-matrices and process them as mini batches respectively. An example is shown in Figure~\ref{fig:blocks}. Given the $n\times n$ similarity matrix $\boldsymbol{S}$, we partition it into four $\frac{n}{2}\times \frac{n}{2}$ sub-matrices. The $n\times n\times 2d$ batch input matrix is also packed accordingly. Then each sub-matrix can be computed through our Bi-LSTM model.

\subsection{Network architecture}
The architecture of the neural network includes two Bi-LSTM layers followed by two fully connected layers. Both Bi-LSTM layers have 512 outputs (256 forward and 256 backward). The first fully connected layer is 64-dimensional with the ReLU activation function. The second layer is 1-dimensional, connected with a sigmoid function to output a similarity score between 0 and 1.

\subsection{Similarity matrix enhancement}
To smooth and denoise the data, we employ the similarity matrix enhancement introduced in \cite{wang2018speaker} with the Gaussian Blur step removed. This operation improves the system performance and the detailed procedure is listed as follows:
%\cite{wang2018speaker} introduces an algorithm for similarity matrix enhancement before spectral clustering. This work is also employed in our system, with the Gaussian Blur step removed. 

\begin{enumerate}[a)]
  \setlength{\itemsep}{1pt}
  \setlength{\parsep}{0pt}
  \setlength{\parskip}{0pt}
  \item Symmetrization: $Y_{i,j} = \max(S_{ij}, S_{j,i})$
  \item Diffusion: $Y \leftarrow YY^\mathrm{T}$
  \item Row-wise max normalization: $S_{ij} = Y_{ij}/\max_{k}Y_{ik}$
\end{enumerate}

\section{Experimental Results}
\label{sec:exp}

\subsection{Dataset}
\label{sec:data}
In our proposed approach, models for extracting speaker embedding vectors including i-vectors and x-vectors are trained on a collection of SRE-databases including SRE 2004, 2005, 2006, 2008 and Switchboard. For the evaluation, we choose NIST SRE 2000 CALLHOME (LDC2001S97) Disk 8, a widely used telephone dataset containing multiple languages with the number of speakers ranging from 2 to 7. There are 500 utterances in total, summing up to about 18 hours. Since our model is supervised, a 5-fold validation is carried out on the evaluation dataset. We split the 500 utterances into five subsets uniformly, and each time one subset is drawn as the evaluation dataset while the other four are fed into our Bi-LSTM model for training. Finally, we combine the 5-fold evaluation results and report system performance. To guarantee fairness, we also conduct the 5-fold validation in PLDA based systems where four training subsets are utilized for whitening PLDA including mean subtraction, full rank PCA mapping and length normalization. 

\subsection{Evaluation metrics}
Speaker diarization systems are usually evaluated by diarization error rate (DER). In order to account for manual annotation error, it is common not to evaluate short collars centered on each speech turn boundary (0.25s on both sides). Overlapped speech regions are excluded. DER consists of three components: false alarm (FA), missed detection (Miss), and speaker confusion, among which FA and Miss are mostly caused by VAD errors. Since an oracle VAD is employed in our implementation, we exclude FA and Miss from our evaluations. Therefore, reported DERs actually correspond to speaker confusion.

\subsection{Training and testing process}
As described in Section~3, our model processes one audio in a batch and the corresponding output is the similarity matrix. We set the maximum matrix size as $400\times 400$ and any larger matrix is partitioned into sub-matrices. In the training process, we reshape both the batch output and the groundtruth ideal similarity matrix into $n^2$ vectors and adopt BCE loss. We rely on stochastic gradient descent for training, with a learning rate initialized at 0.01 and reduced by a factor of 10 every 40 epochs. The whole training process terminates after 100 epochs, 
% Hervé: does the sentence below mean that hyper-parameter tuning is done on the training set as well? Or do you actually 
% Qingjian Lin: yeah, the training outputs are used after system hyper parameters are well tuned.
and then the training outputs are used to tune clustering thresholds. 
In the evaluation process, those thresholds are applied to the test set and DER is used to compare systems.

\subsection{Implementation details}

\noindent\textbf{Speech segmentation}: All experiments share the same segmentation module. A sliding window with duration 1.5s and 750ms overlap is applied on speech regions to generate speaker-homogeneous segments. Each segment is labelled with the most talkative speaker in the central 750ms-long region.

\noindent\textbf{i-vector extraction}: 20-dimensional MFCCs with delta and delta-delta coefficients are extracted to train a 2048-component GMM-UBM model. Supervectors can be computed and projected into 128-dimensional i-vectors through the total variability space $T$. The whole i-vector system is based on the kaldi/egs/callhome\_diarization/v1 scripts~\cite{sell2014speaker,povey2011kaldi}.

\noindent\textbf{x-vector extraction}: 23-dimensional MFCCs are extracted and followed by cepstral mean normalization. Reverberation, noise, music, and babble noises are added to audio files for data augmentation. The whole x-vector system is based on the kaldi/egs/callhome\_diarization/v2 scripts~\cite{povey2011kaldi,sell2018diarization}.

\subsection{Results and discussion}
\label{sec:res}

We carry out the experiments in three stages. First, we construct two baselines based on i-vector and x-vector using PLDA similarity measurement and AHC clustering. Then, we use spectral clustering (SC) instead of AHC as the clustering backend. Finally, we substitute PLDA with our Bi-LSTM model. System fusion is also conducted by weighted sum at the similarity matrix level. Recent works on the same evaluation dataset are compared in Table~\ref{tab:results}.

\begin{table}[htb]
  \caption{DER (\%) on the test set for different systems.}
  \label{tab:results}
  \centering
  \begin{tabular}{llr}
    \toprule
    & \textbf{System architecture}                  & \textbf{DER(\%)}  \\ 
    \midrule
    \multirow{2}{*}{\textbf{Baseline}}            & i-vector + PLDA + AHC                         & 10.42             \\ 
                                                  & x-vector + PLDA + AHC                         & 8.64              \\ \hline
    \multirow{2}{*}{\textbf{SC backend}} & i-vector + PLDA + SC                          & 10.13             \\  
                                                  & x-vector + PLDA + SC                          & 8.05              \\ \hline
    \multirow{3}{*}{\textbf{LSTM scoring}}        & (1) i-vector + LSTM + SC                          & \textbf{8.53}     \\  
                                                  & (2) x-vector + LSTM + SC                          & \textbf{7.73}     \\ 
                                                  & (1+2) system fusion                                 & \textbf{6.63}     \\ \hline
    \multirow{4}{*}{\textbf{Recent works}}        & Wang et al.~\cite{wang2018speaker}            & 12.0              \\ 
                                                  & Sell at al.~\cite{sell2015diarization}        & 11.5              \\
                                                  & Romero et al.~\cite{garcia2017speaker}        & 9.9               \\ 
                                                  & Zhang et al.~\cite{zhang2018fully} (5-fold)    & 7.6               \\ 
    \bottomrule
  \end{tabular}  
\end{table}

As shown in Table~\ref{tab:results}, the proposed LSTM+SC combination beats both PLDA+AHC and PLDA+SC standard approaches, with DER of 8.53\% for i-vector and 7.73\% for x-vector. 
System fusion further pushes the DER to 6.63\%, outperforming all recent diarization systems in the same evaluation dataset
%As shown in Table~\ref{tab:results}, on condition that the same speaker embedding frontend is employed, our single systems report the lowest DERs of 8.53\% and 7.73\% respectively, and system fusion further pushes the DER to 6.63\%, outperforming all recent diarization systems in the same evaluation dataset. 

The superior performance of the proposed similarity measurement is believed to result mainly from the LSTM ability to process sequences. Multi-speaker conversations are usually highly structured and turn-taking behaviors follow hidden laws of statistics over time. PLDA ignores this contextual information, while Bi-LSTMs takes full advantage from forward and backward sequences. 

To support our statements, we conduct Student's t-test on the results of i-vector + PLDA + SC and i-vector + LSTM + SC systems. The 500 test utterances are sorted in increasing duration order and split uniformly into five groups. The first group contains the shortest 100 utterances while the last group contains the longest ones. In each group, we assume utterance DERs follow the normal distribution and carry out t-test analysis. The null ($H_0$) and alternative ($H_1$) hypotheses are set up as:
$$H_0: \text{DER}_{\text{plda}} = \text{DER}_{\text{lstm}}, \quad H_1: \text{DER}_{\text{plda}} \neq \text{DER}_{\text{lstm}}.$$
We set the $p$-value as 0.05 and thus accept $H_0$ if the t-value is in (-$1.96, 1.96$). Results are shown in Table~\ref{tab:win}. $H_0$ is accepted in short utterance groups while rejected in long utterance groups with 95\% confidence. Since $\widehat{\text{DER}}_{\text{lstm}}$ are smaller than $\widehat{\text{DER}}_{\text{plda}}$ on $H_0$-rejected conditions, we can draw conclusions that LSTM performs better than PLDA in longer utterances.
% where $\text{DER}_{\text{plda}}$ and $\text{DER}_{\text{lstm}}$ are average utterance DERs from PLDA and LSTM respectively.

\begin{table}[th]
  \caption{T-test in five groups with different durations.}
  \label{tab:win}
  \centering
  \begin{tabular}{@{\ \ }c@{\ \ \ \ }c@{\ \ \ \ }c@{\ \ \ }c@{\ \ \ }c@{\ \ }}
    \toprule
    sorted utterances       &$\widehat{\text{DER}}_{\text{plda}}$  &$\widehat{\text{DER}}_{\text{lstm}}$  &t-value  &$H_0$  \\
    \midrule
    $1^{st}\sim100^{th}$    &6.6                      &5.5                      &-1.22    &accepted \\
    $101^{th}\sim200^{th}$  &5.7                      &5.3                      &-0.35    &accepted \\
    $201^{th}\sim300^{th}$  &6.1                      &\textbf{3.9}                      &-2.16    &\textbf{rejected} \\
    $301^{th}\sim400^{th}$  &9.2                      &\textbf{7.5}                      &-2.11    &\textbf{rejected} \\
    $401^{th}\sim500^{th}$  &13.9                     &\textbf{11.6}                     &-2.38    &\textbf{rejected} \\
    \bottomrule
  \end{tabular}
\end{table}

Another interesting phenomenon can be observed in Table~\ref{tab:results}. When LSTM-based similarity measurement is applied, the DER gap between i-vector and x-vector is narrowed, from 2.08\% to 0.80\%. It might be because the advantage brought by deep speaker embedding frontend is neutralized by network based similarity measurement backend. Since x-vector was initially brought up for the speaker verification task, its supervised target might be slightly different from that of speaker diarization. One of our future direction is to jointly train speaker embedding frontend and the similarity measurement backend.

\section{Conclusions}
\label{sec:con}
In this paper, we propose a Bi-LSTM model to substitute PLDA in similarity measurement process for the speaker diarization task. Our best system achieves state-of-the-art performance with a 6.63\% DER. Through analysis we point out that the improvement mainly results from sequence perception of the LSTM model on longer recordings.

\section{Acknowledgements}
This research was funded in part by the National Natural Science Foundation of China (61773413), Natural Science Foundation of Guangzhou City (201707010363), Six talent peaks project in Jiangsu Province (JY-074), Science and Technology Program of Guangzhou City (201903010040), ANR through the ODESSA (ANR-15-CE39-0010), PLUMCOT (ANR-16-CE92-0025) and Lenovo. We thank Lin Yang, Xuyang Wang, Junjie Wang, Yulong Liang and yingjie Li from AI lab of Lenovo research who provided insight and expertise that greatly assisted this research.

\bibliographystyle{IEEEtran}
\bibliography{linqj}

\end{document}